\documentclass[twocolumn, prb, superscriptaddress]{revtex4-1}
\usepackage[utf8]{inputenc}
\usepackage{amsmath}
\usepackage{amsfonts}
\usepackage{bm}
\usepackage{amssymb}
\usepackage{graphicx}
\usepackage{ dsfont }

\begin{document}

\author{Andrea De Lucia}
\affiliation{Institute of Physics, Johannes Gutenberg University Mainz, 55128 Mainz, Germany}
\affiliation{Graduate School of Excellence - Materials Science in Mainz, 55128 Mainz, Germany}

\author{Benjamin Krüger}
\affiliation{Institute of Physics, Johannes Gutenberg University Mainz, 55128 Mainz, Germany}

\author{Oleg A. Tretiakov}
\affiliation{Institute for Materials Research, Tohoku University, Sendai 980-8577, Japan}
\affiliation{School of Natural Sciences, Far Eastern Federal University, Vladivostok 690950, Russia}

\author{Mathias Kläui}
\affiliation{Institute of Physics, Johannes Gutenberg University Mainz, 55128 Mainz, Germany}
\affiliation{Graduate School of Excellence - Materials Science in Mainz, 55128 Mainz, Germany}

\title{Multiscale Model Approach for Magnetization Dynamics Simulations}

\begin{abstract}
Simulations of magnetization dynamics in a multiscale environment enable rapid evaluation of the Landau-Lifshitz-Gilbert equation in a mesoscopic sample with nanoscopic accuracy in areas where such accuracy is required. We have developed a multiscale magnetization dynamics simulation approach that can be applied to large systems with spin structures that vary locally on small length scales. To implement this, the conventional micromagnetic simulation framework has been expanded to include a multiscale solving routine. The software selectively simulates different regions of a ferromagnetic sample according to the spin structures located within in order to employ a suitable discretization and use either a micromagnetic or an atomistic model.
To demonstrate the validity of the multiscale approach, we simulate the spin wave transmission across the regions simulated with the two different models and different discretizations. We find that the interface between the regions is fully transparent for spin waves with frequency lower than a certain threshold set by the coarse scale micromagnetic model with no noticeable attenuation due to the interface between the models. As a comparison to exact analytical theory, we show that in a system with Dzyaloshinskii-Moriya interaction leading to spin spiral, the simulated multiscale result is in good quantitative agreement with the analytical calculation.
\end{abstract}
\maketitle

\section{Introduction}
\begin{figure}[hbtp]
\centering
\includegraphics[width=0.95\columnwidth]{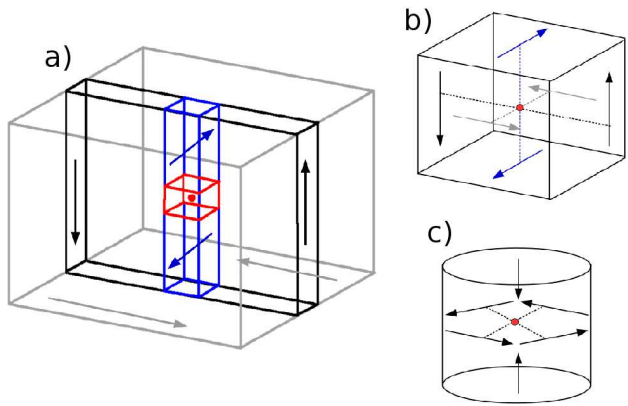}
\caption{(a) Schematic of a magnetization structure with a micromagnetic singularity (Bloch point). The two grey domains are separated by two Bloch walls (black).
The Bloch walls have opposite sense of rotation and are separated by two Néel/Bloch lines (blue).
Between the two Néel/Bloch lines with opposite orientations, a micromagnetic singularity (red) is formed.
A magnification of the red square is shown in (b).
(c) A micromagnetic singularity also occurs during the reversal of a magnetic vortex core. \cite{Hertel, elias-verga}
These diagrams were adapted from Ref.~\onlinecite{tesiBK}. \label{fig:bloch}}
\end{figure}

To model magnetization dynamics, currently two paradigms are commonly used in the field: the micromagnetic model and the Heisenberg spin model.
The micromagnetic model \cite{Aharoni} is ideal when simulating systems with linear dimensions of the order of a few nanometers or larger; since it is a continuous model that is discretized for computational application, its reliability decreases dramatically when simulating magnetic structures exhibiting a large gradient that cannot be resolved by the finite size cells. A textbook example for this scenario is offered by Bloch points \cite{Thiaville} (see Fig.~\ref{fig:bloch}), domain walls and spin waves also belong to this category for particular values of the material parameters.

\begin{figure}[hbtp]
\centering
\includegraphics[width=\columnwidth]{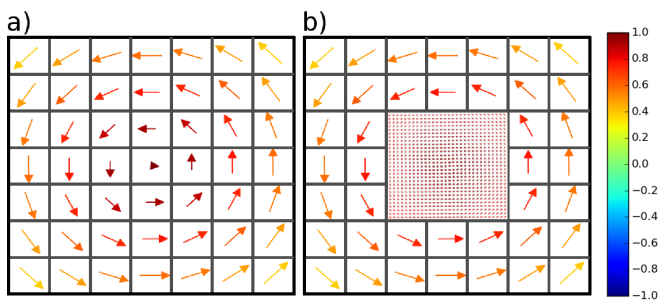}
\caption{Schematic diagram showing the basis of the multiscale model. a) In this example each cell in the vortex core region is simulated in the coarse scale. b) depicts the multiscale simulation, where a small region (central 9 cells) is simulated using the atomistic model, while the rest of the sample is simulated using the micromagnetic model. The color code shows the out of plane component of the magnetization in units of $M_{s}$.\label{fig:geom}}
\end{figure}

The Heisenberg model \cite{Polyakov, Hinzke, Schieback} is a discrete description, where with every atom in the lattice of the ferromagnet a magnetic moment is associated.
Since this is a discrete model, its capability to simulate any magnetic structure is not limited by computational artifacts originating from the discretization of a continuum model, which makes it distinct from micromagnetism.
On the other hand, the Heisenberg model cannot be efficiently used to simulate systems larger than a few nanometers due to the computational time increasing faster than linearly with the number of atoms. \cite{trans, trans2}
In the presented approach (Fig.~\ref{fig:geom}), the entire system is simulated using the micromagnetic model while one or more regions of it containing large gradient structures (e.g. Bloch points), are simulated using the discrete Heisenberg model.
The main obstacle for the development of a combined multiscale technique consists of devising accurate conditions to make the interface between regions on two different scales magnetically smooth, in order to prevent any interface related artifacts.

While in magnetization dynamics, adaptive mesh refinement techniques \cite{Garcia-Cervera, Tako} have been used, none of these employed different models for different scales. One related approach has been proposed, addressing the problem of interfaces between layers of different magnetic materials \cite{Garcia-Sanchez1, Garcia-Sanchez2, Garcia-Sanchez3}. However, the lack of proper interface conditions, in particular the choice of applying a coarse scaled exchange field on the magnetic moments along the interface in the fine scale region, restricts the validity of this approach to the systems with uniform magnetization across the interface. While this shortcoming has been later resolved in Refs.~\onlinecite{Jourdan, Jourdan2}, these approaches were devised to evaluate equilibrium configurations rather than simulating dynamical systems.

One further related approach \cite{Tetramag} employed the finite elements method. It should be noted however that while in this case the atomic lattice in the Heisenberg model can be rendered more accurately, the computational times cannot be dramatically reduced as shown for our finite differences approach in \cite{trans}, making this approach considerably slower. One further multiscale approach \cite{Suess}, devised for a different scale combination than the presented one, proposed to use the micromagnetic model as the fine scale model and the Maxwell equations as the coarse scale model, this is however restricted to systems with slowly varying magnetization. Another work \cite{relativistic} uses special relativity to evaluate a corrective term to the Landau-Lifshitz-Gilbert equation in the case of domain wall motion. In continuum mechanics \cite{tadmor,hertelFD}, multiscale approaches are commonly applied to the investigation of mechanical properties of materials, such as their response to deformations and fractures. However, so far it is unclear whether one can develop such a multiscale model for magnetization dynamics that allows one to carry out valid simulations of systems that cannot be modeled with the currently available approaches. 

In this paper we show the details of the multiscale approach, with a particular focus on the interface conditions that we developed to obtain a smooth interaction between regions on different scales. Finally, demonstrations of the validity for the approach are shown, demonstrating the transmission of spin waves across the scale interface without attenuation, and comparing the simulated ground state for structures exhibiting Dzyaloshinskii-Moriya interaction to the analytical theory.

\section{Method}

\begin{figure}[hbtp]
\centering
\includegraphics[width=\columnwidth]{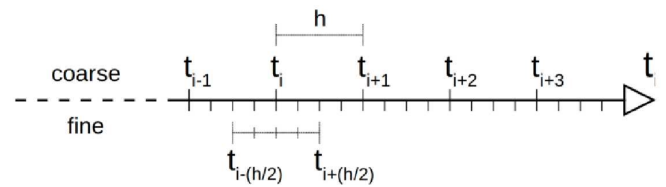}
\caption{Diagram showing the multiscale model in the time domain: after each coarse computational time step the corrections to the effective field in the fine scale region, generated by the coarse one are evaluated, a short series of fine steps centered around the latest coarse step, of length $h$, is executed, then corrections to the coarse scale effective field (generated by the Heisenberg fine scale one) are evaluated. \label{fig:time_coord}}
\end{figure}

The multiscale approach solves the Landau-Lifshitz-Gilbert equation numerically for two different models: the coarse grained micromagnetic model, which simulates the whole sample; and the fine scale model, which is used for magnetic structures that cannot be accurately described by the micromagnetic model, discretizing the magnetization field at atomic resolution and simulating it in the intrinsically discrete Heisenberg spin model.
Our software executes in parallel two independent solving routines, one for each model (it is in principle possible to execute any number of fine scale solving routines), performing one full computational step on the coarse scale one and then a short series of steps on the fine scale one centered around the time coordinate of the coarse one (see Fig.~\ref{fig:time_coord}).

\begin{figure}[hbtp]
\centering
\includegraphics[width=\columnwidth]{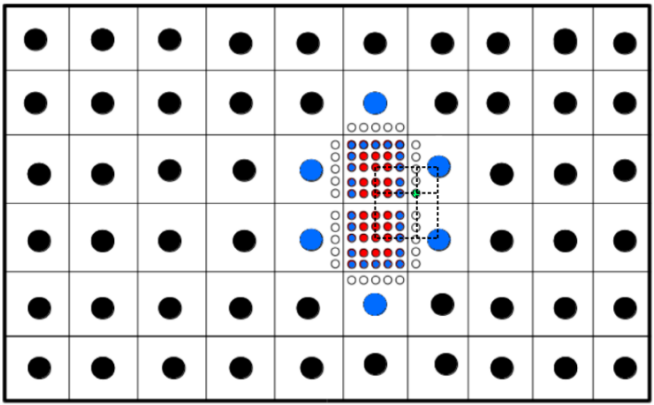}
\caption{The key players in the evaluation of the cross-scale effective field terms: magnetic moments (red), micromagnetic cells (black), interfacial moments and cells (highlited in blue), ghost moments, which are not part of the LLG solving routine (white). The dashed lines show how the ghost moments, and in particular the one marked in green, are evaluated as the bilinear interpolation of fine scale moments and coarse scale magnetization.\label{fig:boundary}}
\end{figure}

The main task towards the development of this technique consisted in modeling the interaction between different regions. This was achieved by applying, after each coarse scale step, a set of magnetic fields designed to approximate the effect of the non-local terms of the effective magnetic field from one region on the other, see Fig.~\ref{fig:boundary}, namely exchange and stray field.
These magnetic fields are designed as follows:
The exchange field, generated by the fine scale magnetic moments closest to the interface ('interfacial moments'), on their 'neighboring' cells in the coarse scale ('interfacial cells') is evaluated by averaging all the interfacial moments inside each coarse scale cell. The average vector is rescaled by the volume $V_a$ of a cell in the atomic lattice, in order to obtain the magnetization $\mathrm{(A/m)}$, rather than the magnetic moment $\mathrm{(Am^2)}$.
A new finite difference mesh, with coarse scale discretization is created and the cells corresponding to the internal surface of the fine scale region are filled with the difference between the magnetization of the same cell in the original coarse mesh and the new vectors. In this way, the linearity of the exchange field with respect to the magnetization is exploited to evaluate a correction to the field, calculated in the micromagnetic formulation, generated by the original coarse scale cells alone.
The corrected exchange field, exerted by the multiscale cell $j$ on the micromagnetic cell $i$ is calculated as:
\begin{equation}
\mathbf{H}_{ex} \left(\mathbf{M}_j \right) = \mathbf{H}_{ex} \left(\mathbf{M}_{int,j} -\mathbf{M}_{j} \right) + \mathbf{H}_{ex} \left(\mathbf{M}_{j} \right).
\end{equation}
Here $\mathbf{M}_{j}$ denotes the magnetization in the cell $j$ in the purely micromagnetic simulation, while $\mathbf{M}_{int,j}$ is defined as:
\begin{equation}
\mathbf{M}_{int,j} = \frac{M_s}{\left|\mu \right| N_{int}}  \sum_{k}^{N_{int}} \bm{\mu}_k = \frac{\sum_{k}^{N_{int}} \bm{\mu}_k}{ V_a N_{int}},
\end{equation} 
where the sum runs over all the magnetic moments $\mu$ located along the interface on the side of cell $j$ that is neighboring cell $i$. This effective field term is evaluated in the micromagnetic model. 
Likewise, to evaluate the exchange field generated by interfacial cells on  interfacial moments, interpolation is employed in order to define a set of new magnetic moments ('ghost moments' \cite{Garcia-Cervera}) to act as first neighbors to the interfacial ones. The exchange field generated by the ghost moments is evaluated in the Heisenberg spin model. A combination of fine scale moments and coarse scale magnetization is used in the interpolation in order to ensure a smooth transition in the magnetic pattern across the interface. This means that each ghost moment results from the interpolation of atomistic and aptly renormalized micromagnetic vectors. The interpolation can be linear, bilinear or quadrilinear according to the dimensionality of the coarse scale mesh. The same techniques, based on the average of interfacial magnetic moments, and the calculation of ghost moments through interpolation across the interface, are employed when evaluating antisymmetric exchange (Dzyaloshinskii-Moriya interaction) across the scale interface.

The stray field contains all the long range contributions to the effective fields. The implementation of this field constitutes one of the main differences between the two models. In both scales the demagnetization tensor formulation was employed,\cite{Newell} as well as the calculation method based on FFT for efficient calculation.\cite{trans} While for the coarse cells the demagnetization tensor describes the interaction between two uniformly magnetized solid rectangles, according to the calculations carried on by Newell \textit{et al.}, \cite{Newell} the demagnetization tensor used for magnetic moments in the fine scale, is defined as:
\begin{equation} \frac{1}{4 \pi} \; \left[ \frac{\mathds{1}}{\left| \mathbf{r}_i - \mathbf{r}_j \right|^3} - 3\frac{\left( \mathbf{r}_i - \mathbf{r}_j \right) \otimes \left( \mathbf{r}_i - \mathbf{r}_j \right)}{\left| \mathbf{r}_i - \mathbf{r}_j \right|^5}\right],
\end{equation}
where $\mathbf{r}_i$ and $\mathbf{r}_j$ are the position of two magnetic moments, $\mathds{1}$ is 3 $\times$ 3 identity matrix, and symbol $\otimes$ denotes the tensor product.

Similarly to the exchange field, the stray field is linear in the magnetization vector and this property is exploited likewise. The correction to the stray field generated in the micromagnetic system by fine scale regions is evaluated using the averaged value of magnetic moments in each cell. 

In order to evaluate the complete demagnetization field acting on the fine scale system, the coarse scale magnetization structure is copied into a new mesh and the cells corresponding to the fine scale region are filled with zero vectors. The stray field generated by this system is evaluated. This technique is employed in order for the field generated by the fine scale region on itself not to be evaluated twice. Since the field has the same discretization as the structure generating it, the result is then interpolated, in order for it to have the discretization of the fine scale mesh. The type of linear interpolation depends, as for the ghost moments, on the dimensionality of the mesh. This is the only case for an effective field term evaluated micromagnetically to be applied on the fine scale region.  This approximation is made necessary by the computational complexity of the algorithm calculating the field, increasing with $N \log(N)$ where $N$ is the number of cells. This dependence is due to the method employed for calculating the demagnetization tensor based on Fast Fourier Transform (FFT). \cite{trans, trans2}

\section{Simulations}

\begin{figure}[hbtp]
\centering
\includegraphics[width=\columnwidth]{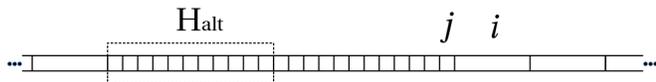}
\caption{Diagram showing the fine scale region of the nanowire and its immediate surroundings. An oscillating magnetic field $H_{\mathrm{alt}}$ is applied to a section of the fine scale region to excite spin waves.
The amplitude of the spin wave is evaluated in the atomistic fine scale cell $j$ and the coarse scale cell $i$ which is described micromagnetically. \label{fig:wire}}
\end{figure}

Having implemented the approach, we run a series of tests as a demonstration of the validity of our model.
The simulated system was a one-dimensional nanowire, $\mathrm{1.8 \, \mu m}$ long with a square $0.3 \times 0.3$ nm$^{2}$ section. The fine scale domain was $\mathrm{90 \, nm}$ long (Fig.~\ref{fig:wire}), the material parameters for this system are those commonly used for permalloy, namely: $M_{s} = 8 \times 10^{5}$ A/m, exchange constant $A = 1.3 \times 10^{-11}$ J/m, and Gilbert damping constant $ \alpha = 0.01$. \cite{micromagnum} For the purpose of efficiency and due to the constraints of the finite difference method, upon which the original software is based, the crystal in the atomistic region is considered to be ordered in a simple cubic lattice with a lattice constant $l=0.3$ nm, comparable to the ones of iron and nickel.

Spin waves of different frequencies $\omega_0$ were excited applying an alternating transversal magnetic field, $H_{\mathrm{alt}}$, to a short ($3 \, \mathrm{nm}$ long) section of the wire.
The magnetization as a function of time was measured on the atomistic moment furthest from the region where $H_{\mathrm{alt}}$ is applied ($\bm{\mu}_j (t)$), and on the neighboring micromagnetic cell ($\mathbf{M}_i (t)$), the transversal component of the two arrays was normalized, and then analyzed using FFT in order to find $\mu_j (\omega)$ and $M_i (\omega)$. Peaks with frequency corresponding to the frequency of $H_{\mathrm{alt}}$ were easily identifiable. The height of such peaks increased linearly with the amplitude of $H_{\mathrm{alt}}$. The peaks, $\mu_j (\omega_0)$ and $M_i (\omega_0)$, were squared and the transmission coefficient $T$ across the interface has been evaluated by calculating the ratio between the two:
\begin{equation}
T(\omega_0) = \frac{\left| \mu_j (\omega_0)\right|^2}{\left| M_i (\omega_0)\right|^2} .
\end{equation}

\begin{figure}[hbtp]
\centering
\includegraphics[width=\columnwidth]{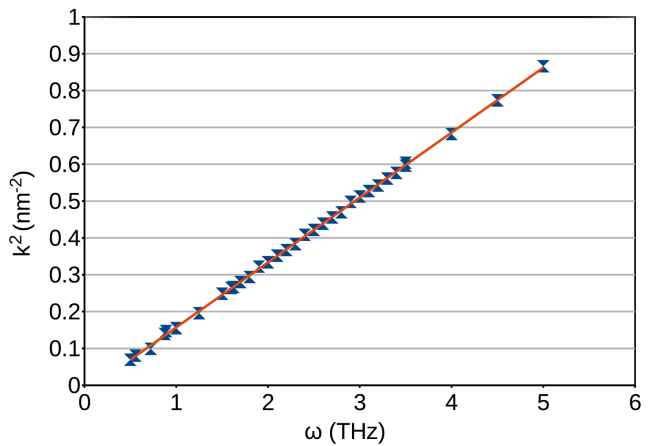}
\caption{Linear regression used to measure the relation between excited wavevector $k$ and the excitation frequency $\omega$. \label{fig:fit}}
\end{figure}

For some values of the frequency, a purely atomistic simulation was performed for comparison and with, the aim of obtaining the relation between frequency and wavelength.
Using FFT in the space domain, the corresponding wavenumber $k$ was measured for each value of the excitation frequency. In particular, such Fourier transforms were evaluated at different time instants and then averaged. Once again peaks were easily identifiable. By means of linear regression (see Fig.~\ref{fig:fit}) the dependence $k^2(\omega)$ was measured and the wavelength corresponding to each value of the excitation frequency was calculated as $\lambda(\omega) = 2 \pi/k(\omega)$.

\section{results}

\begin{figure}[hbtp]
\centering
\includegraphics[width=\columnwidth]{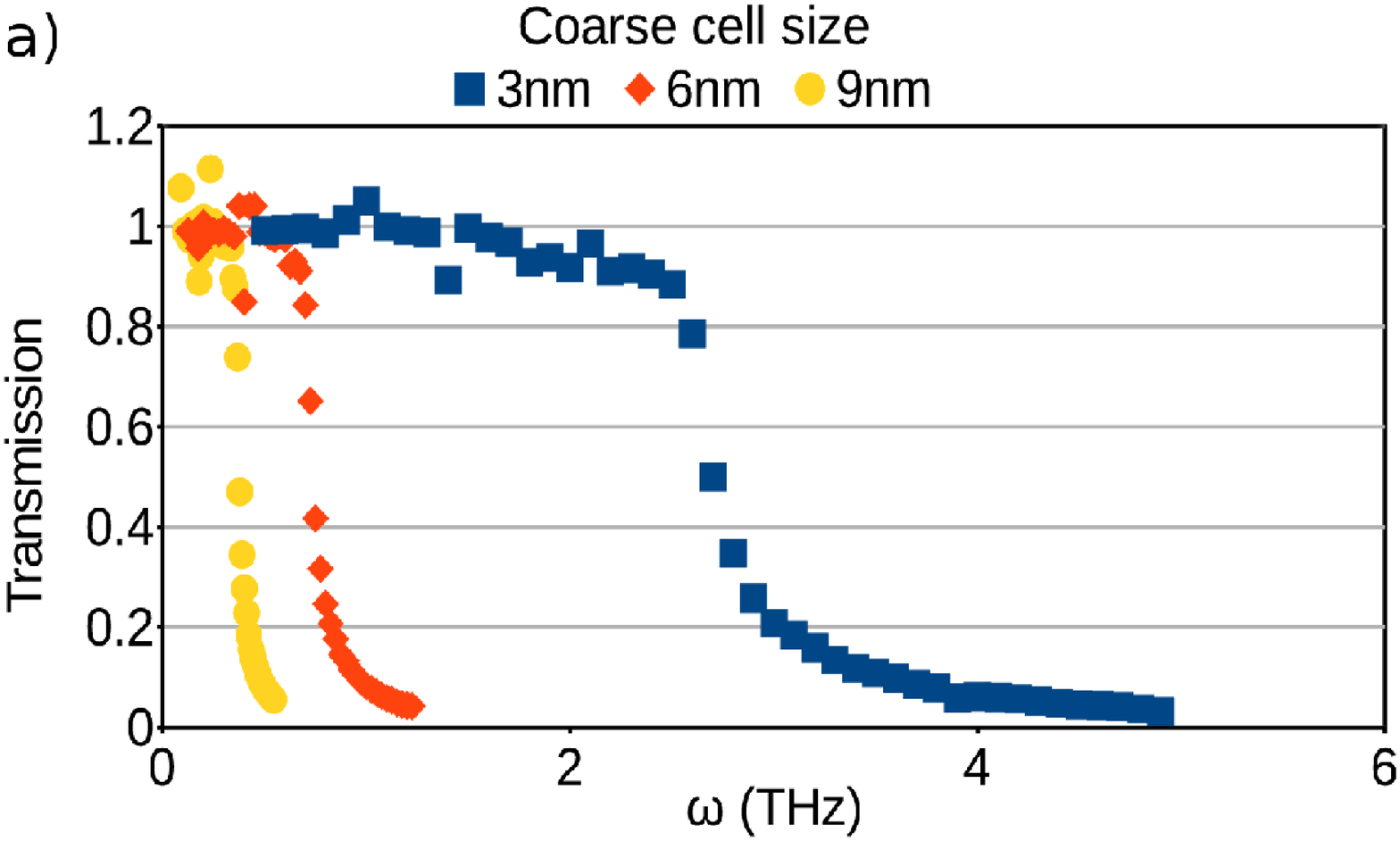}
\includegraphics[width=\columnwidth]{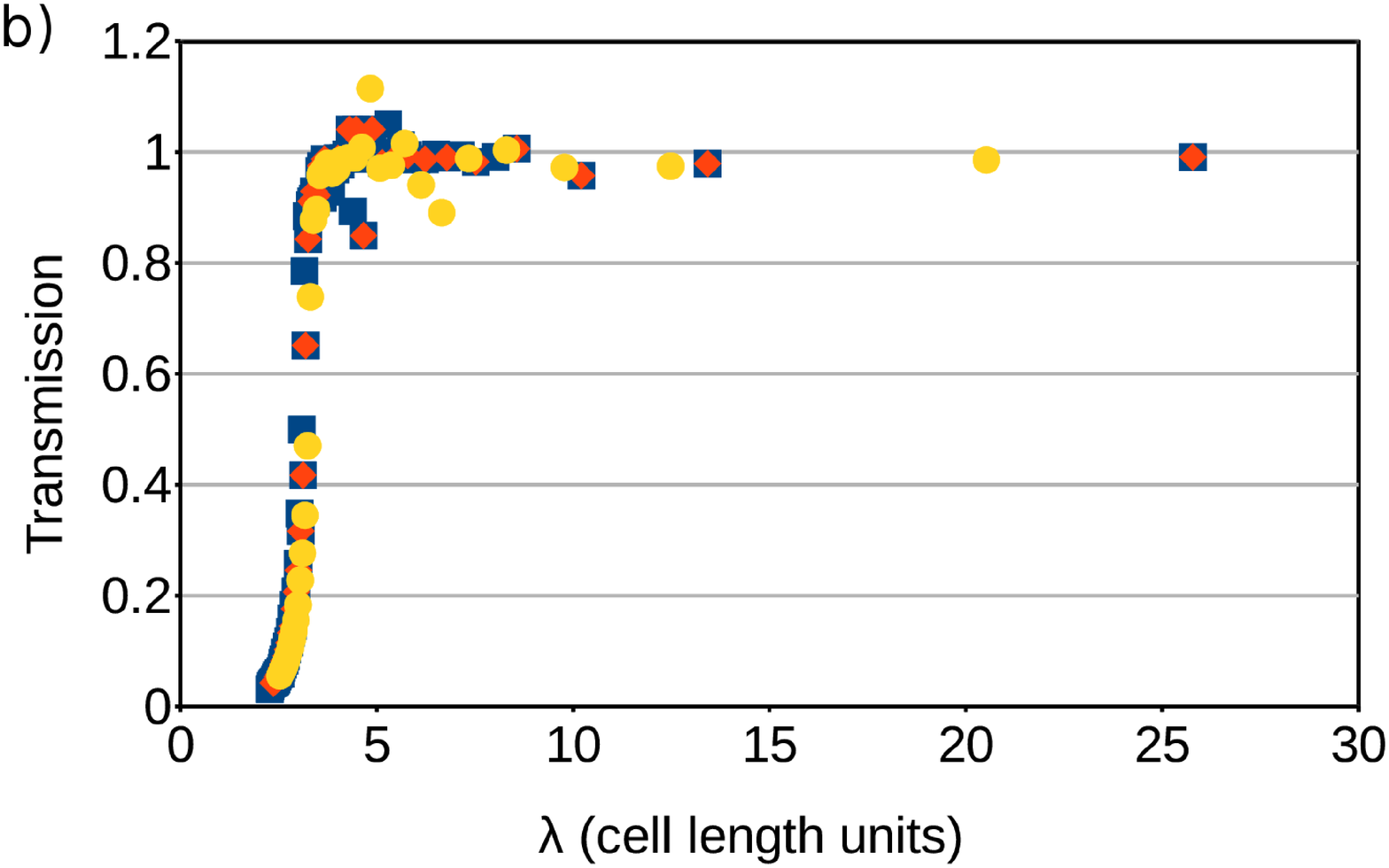}
\caption{The measured transmission for waves excited in the fine scale region with open boundary conditions as a function of their frequency $\omega$ (a) and wavelength $\lambda$ (b). A transmission of 1 ($100 \%$)  for a wide range of wavelengths demonstrates the numerical validity of the model. \label{fig:1}}
\end{figure}

Three sets of simulations were performed, with different lengths of the micromagnetic cells, corresponding to ten, twenty and thirty times $l \;(0.3 \, \mathrm{nm})$. The data show ideal transmission for frequency values smaller than a sharply defined cut-off frequency.
The same data, as a function of the wavelength, show consistently that the transmission drops to zero at a cut-off wavelength corresponding to a specific value of the coarse cell size. This universal behavior can be considered as a limitation of computational micromagnetism, which does not allow one to simulate very short wavelength spin waves without refining the mesh, introducing therefore a dramatic increase in the computation time (Fig.~\ref{fig:1}).

\begin{figure}[hbtp]
\centering
\includegraphics[width=\columnwidth]{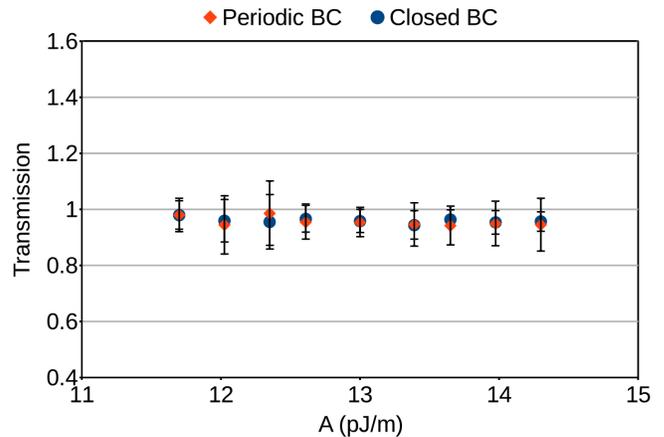}
\caption{The measured average transmission for waves of all possible frequencies excited in the coarse scale region, before entering the fine scale one, with closed and periodic boundary conditions (BC) for different values of the exchange constant $A$. 
The data shown is the result of an average on all the frequencies. Peaks with frequency higher than $3.5 \, \mathrm{THz}$ were not visible in the Fourier transform, underlining the fact that the cutoff is a consequence of the waves not being resolved for the chosen cell-size. The observed transmission of approximately $1$ shows the validity of the method with no artificial attenuation at the interface between the regions where different models are used. \label{fig:2}}
\end{figure}

Since we assume that the frequency cut-off is a consequence of the coarse scale not being able to resolve waves with such a high frequency, we simulated a similar system, this time with the excitation being applied on the coarse scale region only. Here the waves propagate into and then out of the fine scale region and the transmission is measured for waves leaving the fine scale region (Fig.~\ref{fig:2}). The test was repeated using periodic boundary conditions to make sure that the sharp cut-off was not caused by the waves being reflected at the end of the wire. Both tests were then repeated for different values of the exchange constant. 

\begin{figure}[hbtp]
\centering
\includegraphics[width=\columnwidth]{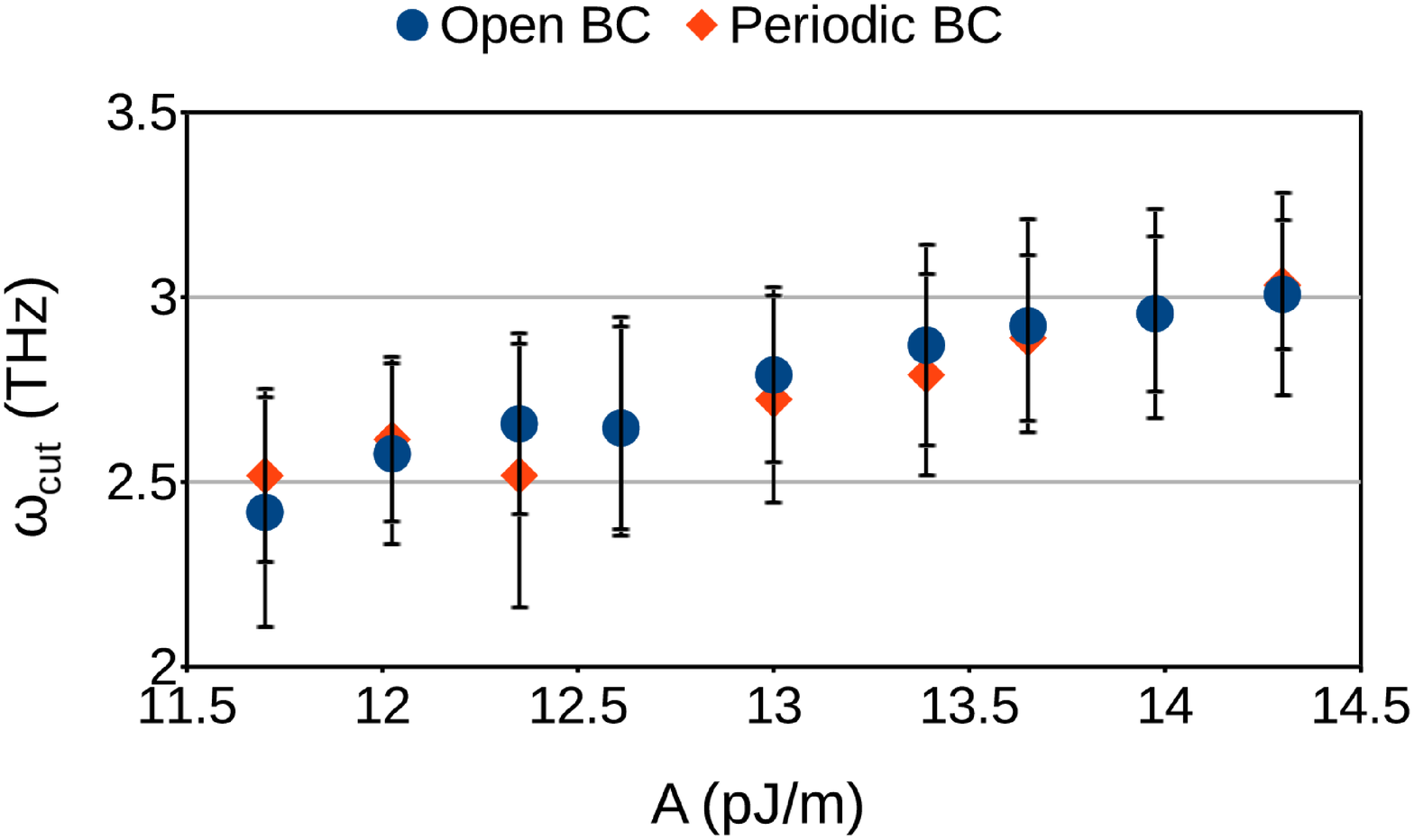}
\caption{The measured cut-off frequency $\omega_{\rm{cut}}$ for waves excited in the fine scale region with closed and periodic boundary conditions (BC) for different values of the exchange constant $A$. 
A cut-off frequency depending on the exchange constant demonstrates that this phenomenon is strictly micromagnetic and is not introduced by the multi-scale approach. \label{fig:3}}
\end{figure}

In order to measure the cut-off frequencies, a linear regression was executed on all the transmission values between $0.1$ and $0.9$, the intersection of this line with the transmission value of $0.5$ was defined as the cut-off frequency.
We assume the cut-off to be a direct consequence of the exchange interaction not being accurately evaluated in the micromagnetic model when the angle in the magnetization between two neighboring cells is too large. The dependence of the cut-off frequency on the exchange constant supports this hypothesis (see Fig.~\ref{fig:3}).

\section{Dzyaloshinskii-Moriya Interaction}

\begin{figure}[hbtp]
\centering
\includegraphics[width=\columnwidth]{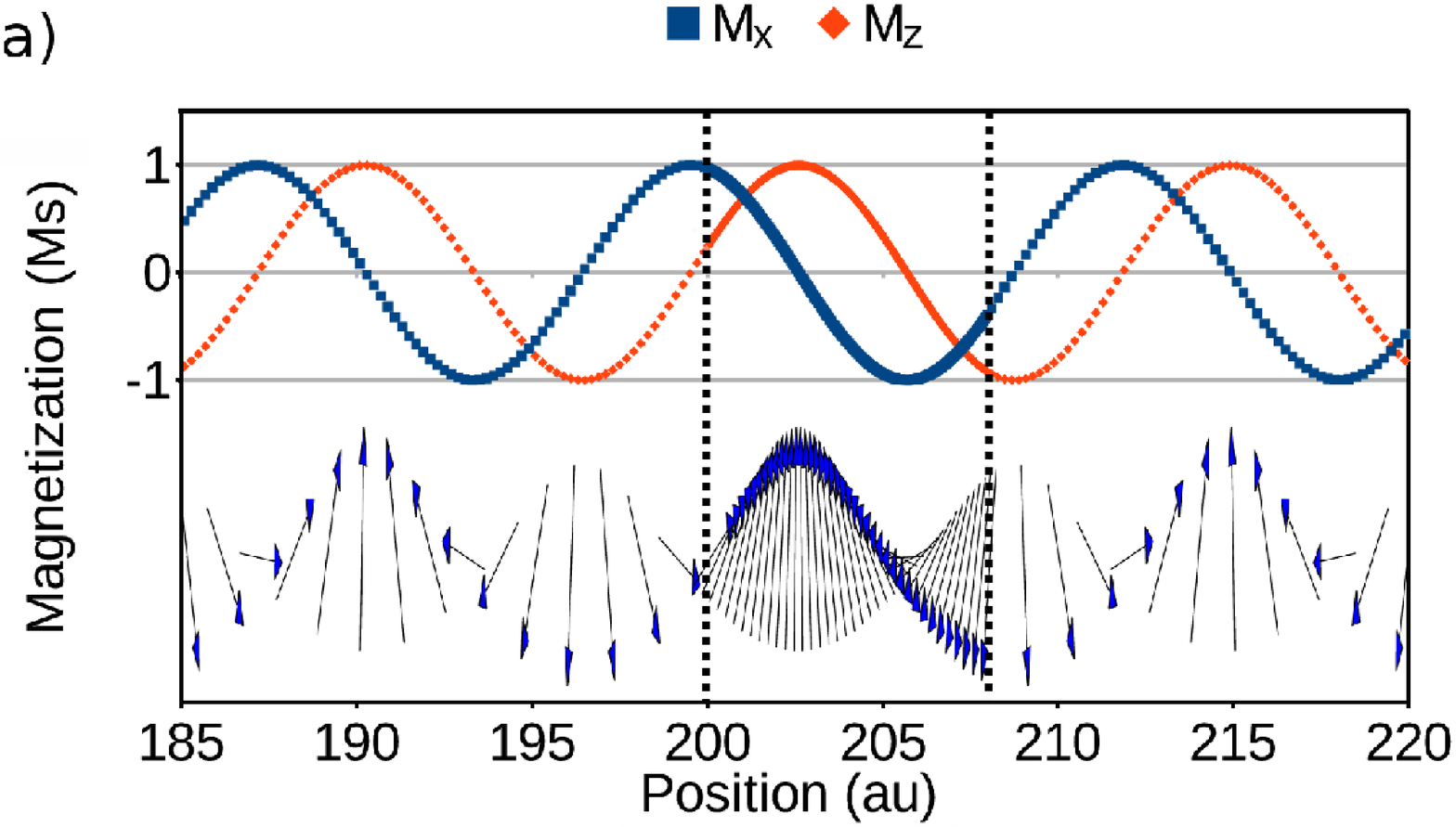}
\includegraphics[width=\columnwidth]{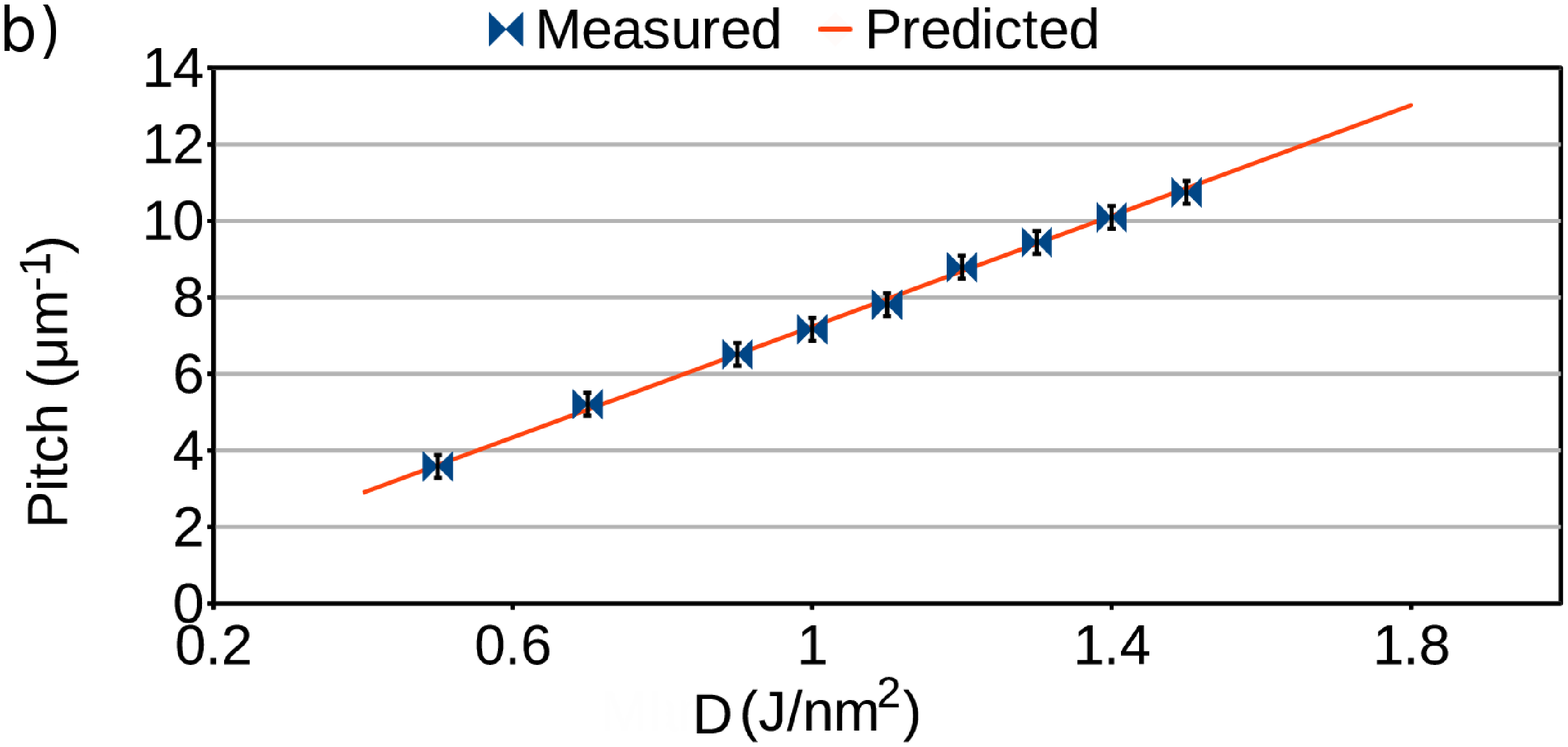}
\caption{ a) The two components of the magnetization for a multiscale DMI helix in the $xz$ plane, showing continuity and consistency of the period in the coarse and fine scale. The dashed lines show the position of the fine scale region. b) The wavenumber of the helix increases linearly with the DMI constant and is consistent with the expected value. \cite{Moriya} \label{fig:DMI}}
\end{figure}

To demonstrate the reliability of the method used to evaluate effective fields across the interface by direct comparison to analytical theory, a system exhibiting antisymmetric exchange\cite{Dzyaloshinskii, Moriya} was simulated. A nanowire, similar in shape to the one used to test spin wave transmission, with the parameters $M_s =1.05 \times 10^6$ A/m, exchange constant $A = 11 \times 10^{11}$ J/m. Different values of $D = \left\vert \bm{D}_{ij} \right\vert$ were used. The vector $\bm{D}_{ij}$ scales the energy density of the Dzyaloshinskii-Moriya interaction (DMI) as calculated in Ref.~[\onlinecite{Moriya}]: $$e_{DMI} = \bm{D}_{ij} \cdot \left(\bm{\mu}_i \times \bm{\mu}_j \right)/\left\vert \mu \right\vert^2. $$ The system was relaxed in a coarse scale simulation, then a fine scale region was applied on a section of the wire and the system was relaxed again. The relaxed state (see Fig.~\ref{fig:DMI}) showing continuity in the helix structure, typical of systems exhibiting DMI, with a pitch in agreement with the predicted\cite{Moriya} value of $D/(4 \pi A)$. The pitch was evaluated from the Fourier transform in the space domain for the two components of the helix, using the data points from both scales and taking the peak value from the Fourier transform. The components of $\mathbf{M}$ evidently have a perfectly sinusoidal shape, see Fig.~\ref{fig:DMI} (a).

\section{Tracking}

A tracking algorithm was devised in order to keep the fine scale region as small as possible, it scans the fine scale region for the position of the structure of interest (SOI), usually the spin structure with large magnetization gradients, and shifts the fine scale region by an integer number of coarse scale cells units, in order to always have the SOI close to its center. When micromagnetic cells previously not part of the fine scale region become included, interpolation is applied in order to fill in the fine scale mesh with magnetic moments that accurately reproduce the coarse scale magnetization and are continuous within and across the scale interface.

\begin{figure}[hbtp]
\centering
\includegraphics[width=\columnwidth]{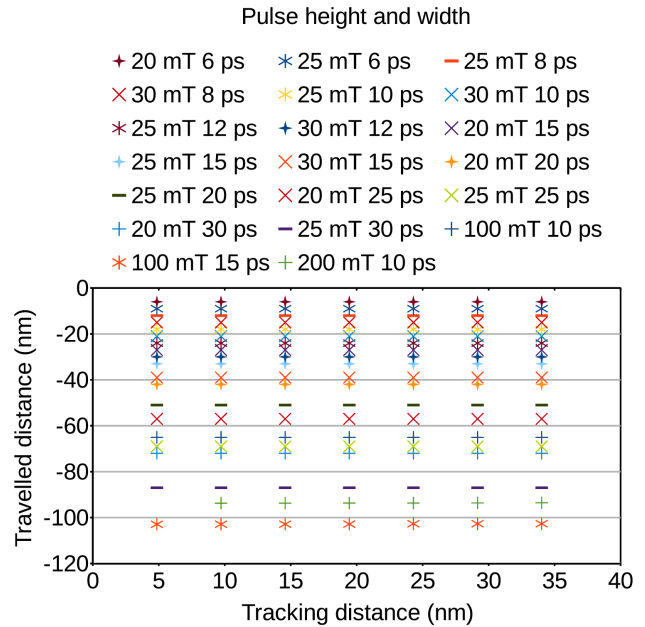}
\caption{Domain wall displacement after the application of a unidirectional Gaussian-shape magnetic field pulse with different values of height and width as a function of the tracking distance. This is the distance traveled by the domain wall before the fine scale region is centered around it. We expect this parameter not to influence the dynamics of the system and the data confirms this assumption.\label{fig:4}}
\end{figure}

To show that the fine scale area can be reliably moved, a test was performed. This test simulated domain wall motion in a nanostrip ($3 \, \mathrm{\mu m} \times 33 \, \mathrm{nm} \times 0.3 \, \mathrm{nm}$) induced by a unidirectional magnetic field. The material parameters of the strip are the same as the nanowire from the previous test with the only exception of Gilbert damping $\alpha = 0.1$. The domain wall is initially in the center of the fine scale region, when the distance from the starting position becomes larger than a certain threshold (tracking distance), the whole fine scale region is shifted, in order to keep it centered. The test was repeated for different tracking distances to show that this process does not influence the dynamics of the system (Fig.~\ref{fig:4}).

\section{Conclusions}

We have presented an innovative methodology to perform magnetization dynamics simulations in the systems which cannot be accurately simulated otherwise. Since some of such systems describe phenomena, including vortex cores switching \cite{Hertel, Schneider, Tchernyshyov, Caputo, Guslienko} and Skyrmion nucleation, \cite{Sampaio, Zhang} are considered to be important problems in spintronics, we deem this methodology a key step to advance this field.
In order to improve the technique and establish multiscale simulations as a valuable tool, its basic features have been described and its limits have been tested.

The transmission data for the spin waves shows that information about magnetic structures in the fine region can cross perfectly the scale interface, thus demonstrating the reliability and numerical validity of our model. A thorough analysis of the cut-off phenomenon found for spin wave transmission shows that in the presence of spin waves with a short wavelength the multiscale approach can be reliably used under the condition that the waves do not leave the fine scale region. Meanwhile, the traditional approach -- a refinement of the whole mesh -- would increase the computational time dramatically.
The simulations including the DMI further show that the method employed for evaluating cross scale interactions ensures continuity between the regions of different scales and yield quantitative agreement with the analytical theory.
Moreover, the domain wall data indicates the reliability of the tracking algorithm and its effectiveness as a method to keep the size of the fine scale regions at a minimum and not introducing artifacts to the simulated results.

As a future direction we propose to analyze the dynamics of magnetic vortex core reversal, a phenomenon that requires a similar approach in order to be accurately simulated. \cite{Hertel} Further research will include magnetic structures such as Skyrmions which are stabilized by DMI and where the nucleation involves Bloch points.
In the long term, there is room for further improvements: generalizing the approach beyond simple cubic lattices in the fine scale region, optimization of the computational routines, extension of this approach to antiferromagnets and nonzero temperatures are some of the examples that will broaden the applicability even further.

\section{Acknowledgements}

A.\,D.\,L. is a recipient of a scholarship through the Excellence Initiative by the Graduate School Materials Science in Mainz (GSC 266), B.\,K. is the recipient of the Carl Zeiss Postdoc Scholarship – Multiskalensimulationen für energiesparende Magnetisierungsmanipulation. The authors acknowledge the support of SpinNet (DAAD Spintronics network, project number 56268455) and the DFG (SFB TRR 173 SPIN+X). O.\,A.\,T. acknowledges support by the Grants-in-Aid for Scientific Research (Grants No. 25800184, No. 25247056 and No. 15H01009) from MEXT, Japan. We are thankful to U. Nowak and D. Hinzke from the University of Konstanz for their help with testing the fine scale model.


\begin{thebibliography}{1}
\bibitem{Aharoni} A. Aharoni, J. Phys. Colloq. \textbf{32}, 966-971 (1971).
\bibitem{Thiaville} A. Thiaville, J.M. Garcia, R. Dittrich, J. Miltat, T. Schrefl, Phys. Rev. B \textbf{67}, 094410 (2003).
\bibitem{Polyakov} A. Polyakov, Phys. Lett. B \textbf{59}, 79-81 (1975).
\bibitem{Hinzke} D. Hinzke and U. Nowak, Phys. Stat. Sol. (a) \textbf{189}, 475-480 (2002).
\bibitem{Schieback} C. Schieback, M. Kläui, U. Nowak, U. Rüdiger, P. Nielaba, Eur. Phys. J. B \textbf{59}, 429-433 (2007).
\bibitem{trans} C. Abert, L. Exl, G. Selke, A. Drews, T. Schrefl, IEEE Trans. Magn. \textbf{48}, 1105-1109 (2012). 
\bibitem{trans2} B. Krüger, G. Selke, A. Drews, D. Pfannkuche, IEEE Trans. Magn. \textbf{49}, 4749-4755 (2013).
\bibitem{Garcia-Cervera} C.J. Garcìa-Cervera and A.M. Roma, IEEE Trans. Magn. \textbf{42}, 1648-1654 (2006). 
\bibitem{Tako} K.M. Tako, T. Schrefl, M.A. Wongsam, R.W. Chantrell, J. Appl. Phys. \textbf{81}, 4082–4083 (1997).
\bibitem{Garcia-Sanchez1} F. Garcia-Sanchez, O. Chubykalo-Fesenko, O. Mryasov, R.W. Chantrell, K.Y. Guslienko, Appl. Phys. Lett. \textbf{87}, 122501 (2005).
\bibitem{Garcia-Sanchez2} F. Garcia-Sanchez, O. Chubykalo-Fesenko, O. Mryasov, R.W. Chantrell, Physica B: Condens. Matter \textbf{372}, 328 (2006).
\bibitem{Garcia-Sanchez3} F. Garcia-Sanchez, O. Chubykalo-Fesenko, O. Mryasov, R.W. Chantrell, K.Y. Guslienko, J. Appl. Phys. \textbf{97}, 10J101 (2005).
\bibitem{Jourdan} T. Jourdan, A. Marty, F. Lancon, Phys. Rev. B \textbf{77}, 224428 (2008).
\bibitem{Jourdan2} T. Jourdan, A. Masseboeuf, A. Marty, F. Lancon, P. Bayle-Guillemaud, J. Appl. Phys. \textbf{106}, 073913 (2009).
\bibitem{Tetramag} C. Andreas, A. Kakay, R. Hertel, Phys. Rev. B \textbf{89}, 134403 (2014).
\bibitem{Suess} F. Bruckner, M. Feischl, T. Führer, P. Goldenits, M. Page, D. Praetorius, M. Ruggeri, D. Suess, Math. Models Methods Appl. Sci. \textbf{24}, 2627 (2014).
\bibitem{relativistic} P. Weinberger, E.Y. Vedmedenko, R. Wieser, R. Wiesendanger, Philos. Mag. \textbf{91}, 2248-2262 (2011).
\bibitem{tadmor} R.E. Miller and E.B. Tadmor, Model. Simulat. Mater. Sci. Eng. \textbf{17}, 053001 (2009).
\bibitem{hertelFD} C. Hertel, M. Schümichen, S. Löbig, J. Fröhlich, J. Lang, Theor. Comp. Fluid. Dyn. \textbf{27}, 817-841 (2012).
\bibitem{Hertel} R. Hertel, S. Gliga, M. Fahnle, C.M. Schneider, Phys. Rev. Lett. \textbf{98}, 117201 (2007).
\bibitem{elias-verga} R.G. Elías and A. Verga, Eur. Phys. J. B \textbf{82}, 159 (2011).
\bibitem{tesiBK} B. Krüger, Doctoral dissertation available at: \\ \url{ediss.sub.uni-hamburg.de/volltexte/2012/5887/pdf/Dissertation.pdf} (2011).
\bibitem{Newell} A.J. Newell, W. Williams, D.J. Dunlop, J. Geophys. Res. \textbf{98}, 9551-9555 (1993).
\bibitem{micromagnum} The MicroMagnum software is available at: \\ \url{micromagnum.informatik.uni-hamburg.de/} \,.
\bibitem{Schneider} R. Hertel and C.M. Schneider, Phys. Rev. Lett. \textbf{97}, 177202 (2006).
\bibitem{Tchernyshyov} O. A. Tretiakov and O. Tchernyshyov, Phys. Rev. B \textbf{75}, 012408 (2007).
\bibitem{Caputo} J.G. Caputo, Y. Gaididei, F.G. Mertens, D.D. Sheka, Phys. Rev. Lett. \textbf{98}, 056604 (2007).
\bibitem{Guslienko} K.Y. Guslienko, K.S. Lee, S.K. Kim, Phys. Rev. Lett. \textbf{100}, 027203 (2008).
\bibitem{Sampaio} J. Sampaio, V. Cros, S. Rohart, A. Thiaville, A. Fert, Nat. Nano. \textbf{8}, 839–844 (2013).
\bibitem{Zhang} X. Zhang, M. Ezawa, Y. Zhou, Sci. Rep. \textbf{5}, 9400 (2015).
\bibitem{Dzyaloshinskii} I. Dzyaloshinskii, \textit{A Thermodynamic Theory of “Weak” Ferromagnetism of Antiferromagnetics}. J. Phys. Chem. Solids \textbf{4}, 241-255 (1958)
\bibitem{Moriya} T. Moriya, \textit{Anisotropic Superexchange Interaction and Weak Ferromagnetism}. Phys.Rev. \textbf{120}, 91 (1960).

\end{thebibliography}
\end{document}